\documentstyle[epsf]{mn2e}

\title[An absorption origin for the soft excess in Seyfert 1 AGN]
{An absorption origin for the soft excess in Seyfert 1 AGN}

\author[M.Middleton, C. Done, M. Gierli{\'n}ski, ]
{Matthew Middleton$^1$, Chris Done$^1$ and Marek~Gierli\'nski$^{1,2}$\\
$^1$Department of Physics, University of Durham, South Road, Durham
DH1 3LE,
UK\\
$^2$Obserwatorium Astronomiczne Uniwersytetu Jagiello{\'n}skiego,
30-244
Krak{\'o}w, Orla 171, Poland\\
}

\pagerange{\pageref{firstpage}--\pageref{lastpage}} \pubyear{2006}

\begin{document}

\topmargin = -0.5cm

\maketitle

\label{firstpage}

\begin{abstract}

The soft excess seen in the X-ray spectra of many AGN can be well modelled
by reflection from a partially ionised accretion disc. However, this often
requires extreme parameters, both in terms of the underlying spacetime and
the reflection geometry, and requires that the disc is far from
hydrostatic equilibrium.  An alternative model uses similarly partially
ionised, velocity smeared material but from an accretion disc wind seen in
absorption. We explicitly compare these two models for the origin of the
soft excess using XMM-Newton data for PG QSOs and Narrow Line Seyfert 1's.
We find that while reflection and absorption give comparably good fits to
the data, the absorption model allows a much clearer correspondence with
the stellar mass black holes. All the objects are high mass accretion rate
AGN, so should be analogous to the high/soft and very high states in black
hole binaries. The intrinsic spectral indices derived from the absorption
model are all consistent with a one-to-one mapping between spectral state
and AGN type, with the NLS1's having softer spectra corresponding to the
very high state, while the broad line AGN have Gamma~2 as expected for the
high/soft state.  By contrast, a few AGN have intrinsically hard spectra
with the reflection model. While this supports an absorption
interpretation of the soft excess, we note that the required Gaussian
velocity dispersion of $\geq 0.2c$ (corresponding to an outflow velocity
$\ge 0.4c$) is too fast for a radiatively driven accretion disc wind and
instead requires that the material is entrained in a magnetic outflow
(jet). We also use the simultaneous OM data to derive the ratio of disc to
total accretion power which is another tracer of spectral state in X-ray
binaries. This does not always show that the disc in NLS1's contributes
less than 80 per cent of the total power, as expected for a very high
state. We suggest that this is an artifact of the standard disc models
used to determine the disc luminosity in our fits. The disc seen in the
very high state of black hole binaries is often observed to be distorted
from the standard shape, and a similar effect in NLS1's could recover the
one-to-one mapping between black hole binary spectral state and AGN type.

\end{abstract}

\section{Introduction}

X-ray spectra and variability provide important evidence to suggest
that accretion systems of AGN are analogous to those of their smaller
counterparts, the black hole binary (BHB) systems. Both show two major
components in their high energy spectra, the accretion disc, together
with a tail of emission extending to higher energies.  The accretion
disc temperature should scale with both mass ($m=M/M_\odot$) and mass
accretion rate (hereafter parameterised as $L/L_{Edd}$) as $T\sim 10^7
(L/L_{Edd})^{1/4} m^{-1/4}$~K (Shakura \& Sunyaev 1973). This
component peaks in the soft X--ray range for bright BHB, but 
correspondingly large $L/L_{Edd}$ AGN have discs which peak in the
UV--optical region of the spectrum. This complicates a detailed
comparison of the two systems. In BHB the X--ray spectra contain
contributions from both the disc and tail, whereas in AGN only the
tail should contribute. This is particularly important as BHB show a
complex range of {\em spectral states}, which can be roughly
determined by the luminosity ratio between the disc and tail, together
with the shape of the tail (see e.g. Remillard \& McClintock 2006). In
BHB these can all be (more or less) determined simply from a single
X--ray spectrum, whereas to constrain the disc emission in AGN
requires simultaneous UV--optical as well as X--ray data. This is now
easily obtainable with the launch of the XMM-Newton satellite, which
carries the Optical Monitor telescope with optical/UV filters, as well
as the sensitive EPIC X--ray CCD's.

In BHB the disc and high energy tail are well known to have different
variability properties on short timescales, where the disc is rather
constant while the high energy tail is not (e.g Churazov Gilfanov \&
Revnivtsev 2001). The effects of the stable accretion disc diluting
the observed X--ray variability is important in BHB, but when this
effect is corrected for then the AGN and BHB show very similar
variability power spectra (Done \& Gierli{\'n}ski 2005), with
characteristic frequencies simply scaling with mass for a given
$L/L_{Edd}$ (McHardy et al 2006). This points to the same underlying
physical mechanisms for forming the tail in both systems, strongly
supporting models in which the accretion processes are the same.

However, despite this overall correspondence between the two systems,
there are clear discrepancies which remain unexplained. Firstly there
is the `soft excess' seen ubiquitously in high $L/L_{Edd}$ AGN. This
is a smooth rise above the extrapolation of the higher energy continuum below $\sim
$~1~keV. This energy is too high to be explained by the standard
accretion disk model above, and alternative continuum models such as
Comptonization of the disc (e.g. Czerny \& Elvis 1987; Page et al
2004) or a super-Eddington slim disk (e.g. Abramowicz et al. 1988;
Tanaka, Boller \& Gallo 2005) have problems in explaining why this
component is always seen at the same energy in AGN spanning a large
range in mass (Gierli{\'n}ski \& Done 2004: hereafter GD04), and why this
has no obvious counterpart in the BHB (GD04; Sobolewska \& Done 2007).
Secondly, some of these high $L/L_{Edd}$ have apparently flat
2--10~keV spectra. BHB at correspondingly high $L/L_{Edd}$ normally
have rather steep tails to high energies, with photon index $\Gamma >
2$ when characterised by a power law (e.g. the review by Remillard \&
McClintock 2006).

Both these problems can be addressed if there is substantial
distortion of the AGN spectra by atomic processes. This rather
naturally gives rise to a fixed energy for the soft excess from the
dramatic increase in opacity in partially ionised material at $\sim$
0.7~keV provided by O {\sc VII}/O {\sc VIII} and the Fe M shell
unresolved transition array (UTA). The observed smoothness of the soft
excess then requires that this partially ionised material has strong
velocity gradients to remove any lines or edges by Doppler smearing.
This material can affect the spectrum in two rather different
geometrical configurations, either in optically thick material seen in
reflection (e.g. Crummy et al. 2006) or optically thin material seen
in absorption (e.g. Gierli{\'n}ski \& Done 2004; Chevallier et
al. 2006), but in both models the enhanced opacity above 0.7~keV can
lead to an apparently flat 2--10~keV spectrum even when the intrinsic
emission is steep.

This all makes a compelling case for there to be substantial amounts
of partially ionised, Doppler smeared material in AGN. Even if the
same material is present in BHB it would not be easily seen as the
higher temperature accretion disc will result in complete
photoionization of Oxygen, so will not produce the strong opacity jump
at 0.7~keV. In which case the question becomes whether this material
is seen in reflection or absorption. This has additional importance as
the difference between absorption and reflection mechanisms implies
not only differing physical geometries but also a difference in the
inferred properties of the spacetime (Gierli\'nski \& Done 2006).

In the reflection models, the strength of the soft excess is limited
if the source is isotropic. The maximum reflected continuum below
0.7~keV equals the incident continuum, and even including line
emission results in a soft excess of no more than a factor of 2-3
above the extrapolated 2--10~keV continuum (Sobolewska \& Done 2007).
This is exceeded by many high $L/L_{Edd}$ AGN, dramatically so by 
a few objects. Thus the
reflection geometry also requires anisotropic illumination from
either gravitational light bending or a corrugated disc surface which
acts to hide a large proportion of the incident flux from the line of
sight (Fabian et al. 2002a; Miniutti et al 2003; Miniutti \& Fabian
2004). The problem becomes even more marked if the reflecting disc
material is in hydrostatic equilibrium, as the details of the pressure
balance means that there is only a small optical depth in the disc
photosphere where Oxygen can be partially ionised. Thus these models
give much smaller soft excess than a constant density reflecting slab
(Done \& Nayakshin 2007).

Instead, the absorption model sets no such restrictive limits on the
size of the soft excess, so does not require any geometry change to
produce the strongest observed soft excesses (Sobolewska \& Done
2007). Additionally, and in stark contrast to the reflection models,
any pressure balance condition rather naturally produces the required
(now much smaller) optical depth of partially ionised Oxygen
(Chevallier et al 2006).

While this all supports the absorption model, these arguments are
indirect, relying on physical plausibility. A clear observational test
between the two models would be more convincing. However, detailed studies
of individual objects have shown that even good quality XMM-Newton data in
the 0.3--10~keV bandpass cannot distinguish between reflection and
absorption model for the origin of the spectral features (Chevallier et
al. 2006; Sobolewska \& Done 2007). Similarly, the 0.3--10~keV variability
rms spectra of individual objects can also be equally well matched by both
models (Ponti et al. 2006; Gierli\'nski \& Done 2006). This is the case
even for the canonical reflection dominated, broad line source,
MCG-6-30-15, for which the 0.3-10~keV spectrum and r.m.s. variability can
be well fit by both reflection and absorption models (Done 2007;
Gierli\'nski \& Done 2006). Even extending the bandpass to 40~keV (such as
with the Suzaku HXD: Miniutti et al 2007) does not break the model
degeneracy of the spectra for this object (Done 2007), though the fact that the
variable component can be well fit by a simple absorbed power law
is plainly consistent with the reflection modelling (Miniutti et
al 2007).

Here instead we analyse the spectra of a large sample of PG quasars
and Narrow Line Seyfert 1 galaxies observed with the absorption
models. This sample has previously been studied by Crummy et al
(2006), who fit the smeared reflection model so allowing a direct
comparison of fit quality over a larger number of objects, with a
large spread in spectral shape. We use XMM-Newton data so we can
directly constrain the disc simultaneously with the high energy tail,
and so directly test whether these do indeed correspond to the BHB
high $L/L_{Edd}$ spectra states, and identify where the anomalous soft
excess and hard 2--10~keV spectra arise.

\begin{table*}
\begin{center}
\begin{minipage}{150mm} 
\bigskip

\begin{tabular}{lc|l|l||c|c|c|c}
  \hline
  Source &  z & logM/M$_{\odot}$ &  L/L$_{Edd}$ &Galactic column & $\Gamma_{2-10keV}$& Soft Excess  &Obs. Date\\
  &&&&(x10$^{20}$cm${^-2}$)&&at 0.5keV&\\
 \hline
 
 PG 0003+199  & 0.025 & 7.07 & 0.62 & 3.99 & 2.216 $_{-0.016}^{+0.016}$ & 2.680 & 2000-12-25\\
 PG 0050+124  & 0.061 & 7.13 & 1.81 & 4.99 & 2.284 $_{-0.024}^{+0.025}$& 1.008 & 2002-06-22\\
 PG 0157+001  & 0.163 & 8.00 & 0.54 & 2.59 &  2.117 $_{-0.139}^{+0.145}$ & 1.712 &2000-07-29\\
 PG 0844+349  & 0.064 & 7.66 & 0.41 & 2.98 & 2.126 $_{-0.034}^{+0.034}$& 2.621 &2000-11-04\\
 PG 0947+396  & 0.206 & 8.46 & 0.14 & 1.57 & 1.843 $_{-0.057}^{+0.058}$& 2.682 &2001-11-03\\
 PG 0953+414  & 0.239 & 8.52 & 0.58 & 1.14 & 1.974 $_{-0.053}^{+0.053}$& 2.512 &2001-11-22\\
 PG 1048+342  & 0.167 & 8.14 & 0.25 & 1.83 & 1.804 $_{-0.055}^{+0.056}$& 2.367 & 2002-05-13\\
 PG 1115+407  & 0.154 & 7.44 & 0.82 & 1.91 & 2.133 $_{-0.073}^{+0.074}$& 2.800 & 2002-05-17\\
 PG 1116+215  & 0.177 & 8.41 & 0.74 & 1.28 & 2.076 $_{-0.069}^{+0.070}$& 2.356 & 2001-12-02\\
 PG 1202+281  & 0.165 & 8.37 & 0.11 & 1.67 & 1.657 $_{-0.044}^{+0.045}$& 2.543 & 2002-05-30\\
 PG 1211+143  & 0.080 & 7.81 & 1.14 & 2.75 & 1.765 $_{-0.024}^{+0.024}$& 2.961 & 2001-06-15\\
 PG 1244+026  & 0.048 & 6.24 & 3.97 & 1.75 & 2.486 $_{-0.084}^{+0.086}$& 3.609 & 2001-06-17\\
 PG 1307+085  & 0.155 & 8.50 & 0.24 & 2.11 & 1.494 $_{-0.071}^{+0.072}$& 2.454 & 2002-06-13\\
 PG 1309+355  & 0.184 & 8.20 & 0.33 & 1.03 & 1.755 $_{-0.084}^{+0.085}$& 1.614 & 2002-06-10\\
 PG 1322+659  & 0.168 & 7.74 & 0.81 & 2.01 & 2.156 $_{-0.096}^{+0.099}$& 2.925 & 2002-05-11\\
 PG 1352+183  & 0.158 & 8.20 & 0.29 & 2.05 & 1.926 $_{-0.087}^{+0.089}$& 4.075 & 2002-07-20\\
 PG 1402+261  & 0.164 & 7.76 & 1.24 & 1.47 & 2.069 $_{-0.079}^{+0.080}$& 4.227 & 2002-01-27\\
 PG 1404+226  & 0.098 & 6.65 & 1.90 & 2.14 & 1.002 $_{-0.455}^{+0.548}$& 77.61 & 2001-06-18\\
 PG 1427+480  & 0.221 & 7.86 & 0.57 & 1.88 & 1.922 $_{-0.054}^{+0.055}$& 2.363 & 2002-05-31\\
 PG 1440+356  & 0.077 & 7.28 & 1.07 & 1.03 & 2.276 $_{-0.050}^{+0.051}$& 3.391 & 2001-12-23\\
 PG 1444+407  & 0.267 & 8.17 & 0.71 & 1.25 & 2.107 $_{-0.126}^{+0.139}$& 3.752 & 2002-08-11\\
 PG 1501+106  & 0.036 & 8.23 & 0.12 & 2.34 & 1.940 $_{-0.025}^{+0.025}$& 2.769 & 2001-01-14\\

 \hline

\end{tabular}
\caption{Source properties of the PG quasars analysed here. Mass and
luminosity are taken from Gierl\'inski \& Done (2004), based on 
Boroson (2002). The Soft Excess strength is the ratio of the source 
data to the extrapolated 2--10~keV power law at 0.5keV.}

\end{minipage}

\end{center}
\end{table*}

\section{data}

We use the same source sample of high mass accretion rate AGN (PG
Quasars and NLS1's) as Crummy et al (2006).  Those which overlapped the
sample of GD04 had spectra extracted from the XMM-\textit{Newton}
database using SAS 5.4 software, selecting only X-ray events
corresponding to patterns 0-12 (MOS) and 0-4 (PN), while those not
included in the GD04 sample were extracted using SAS 7.0 with appropriate checks for consistency in extraction method and ouput performed. Source and
background spectra were extracted from circular regions of 45 arcsec
in radius around the object and in an offset position clear of other
sources respectively.  XMM-\textit{Newton} Optical Monitor (OM) data
was extracted using \textit{omichain}, and the extracted instrumental
magnitudes are converted to flux points, calibrated with respect to
Vega in each filter, with 
calibration values and corresponding fluxes are
provided by the XMM-\textit{Newton} Science Operations
Centre.\footnotemark
\footnotetext{http://xmm.vilspa.esa.es/sas/documentation/watchout/uvflux.shtml}

\begin{table*}
\begin{center}
\begin{minipage}{160mm} 
\bigskip

\begin{tabular}{lc|l|l||c|c|c|c}
  \hline
  Source &  z & logM/M$_{\odot}$ &  L/L$_{Edd}$ & Galactic column  & $\Gamma_{2-10keV}$& Soft Excess  &Obs. Date\\
&&&&(x10$^{20}$cm${^-2}$)&&at 0.5keV&\\
  \hline

NGC 4051  & 0.002 & 6.13 $^{f}$& 0.31 $^{a}$ & 1.32 & 1.245 $_{-0.048}^{+0.048}$& 3.601 &2001-05-16\\
IRAS 13349+2438 & 0.108 & 8.90 $^{d}$ & 0.24$^{i}$& 1.16 & 1.966 $_{-0.043}^{+0.044}$& 3.634 &2000-06-20\\
ARK 564 & 0.025 & 6.46 $^{e}$& 2.75  $^{a}$  & 6.40 & 2.501 $_{-0.027}^{+0.027}$& 2.257 &2000-06-17\\
MRK 1044 & 0.016 & 6.23 $^{e}$& 1.25  $^{a}$ & 3.55 & 1.937 $_{-0.061}^{+0.062}$& 3.472 &2002-07-23\\
E 1346+266 & 0.915 & 8.30 $^{b}$ & 1.40 $^{b}$ & 1.18 & 2.299 $_{-0.239}^{+0.290}$& 2.612 &2003-01-13\\
MRK 0359 & 0.017 & 6.23 $^{e}$& 1.79  $^{a}$  & 4.80 & 1.765 $_{-0.062}^{+0.063}$& 2.312 &2000-07-09\\
PHL 1092 & 0.396 & 8.20 $^{b}$ & approx 1.00 $^{b}$  & 4.07 & 1.979 $_{-0.443}^{+0.355}$& 18.621 &2003-01-18\\
RE J1034+396 & 0.042 & 6.45 $^{a}$&  0.28 $^{a}$ & 1.47 & 2.290 $_{-0.246}^{+0.256}$& 7.325 &2002-05-01\\
PKS 0558-504 & 0.137 & 7.65 $^{c}$ & 1.72 $^{g}$ & 4.38 & 2.255 $_{-0.052}^{+0.053}$& 2.344 &2000-05-24\\
MRK 0766 & 0.013 & 6.63 $^{e}$& 0.47 $^{a}$ & 1.71 & 1.917 $_{-0.028}^{+0.028}$& 1.624 &2000-05-20\\
MRK 0586 & 0.155& 7.86 $^{f}$& 3.37 $^{h}$ & 3.51 & 2.245 $_{-0.082}^{+0.083}$& 2.659 &2002-07-23\\
TON S180 & 0.062 & 7.06 $^{f}$& 2.98  $^{a}$  & 1.55 & 2.286 $_{-0.033}^{+0.034}$& 2.461 &2000-12-14\\

\hline

\end{tabular}
\caption{As for Table 1 but for the non-PG QSOs used in the extended
AGN sample of Crummy et al (2006). References are to $^{a}$ Hao et
al. (2005), $^{b}$ Dasgupta et al. (2004), $^{c}$
Brinkmann et al. (2004), $^{d}$ Blustin et al. (2005), $^{e}$ 
Wang \& Lu (2001), $^{f}$ Woo \& Urry (2002), $^{g}$ 
O'Brien et al. (2001), $^{h}$ Bian \& Zhao (2003),
$^{i}$ Beichman et al. (1986).}

\end{minipage}

\end{center}
\end{table*}

\subsection{Sample properties}

\begin{figure}
\begin{center}
\leavevmode \epsfxsize=7cm \epsfbox{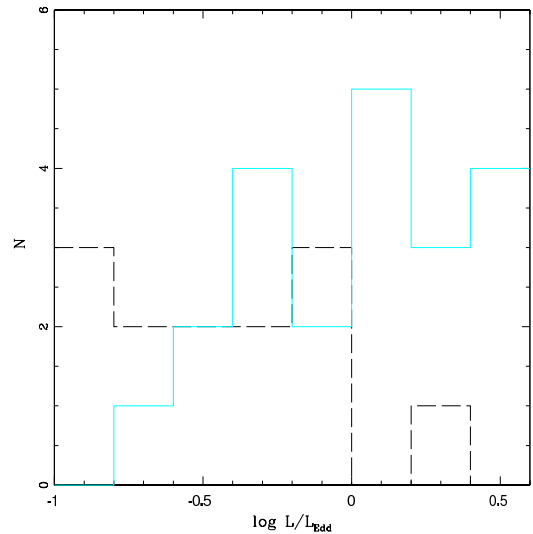}
\end{center}
\caption{Histogram showing the distribution of luminosities across the
sample. The NLS1s (cyan) clearly dominate the high luminosity end, with
broad line AGN (black) having generally lower L/L$_{bol}$.}
\label{fig:hist_Lall}
\end{figure}

The Crummy et al (2006) sample contains 21 NLS1 and 13 broad line
AGN. The distribution of luminosities (from tables 1 \& 2) are shown
as a histogram in Fig.~\ref{fig:hist_Lall} and clearly show that the
NLS1 AGN are the major contributors at the high luminosity end. A
$\chi^2$ test shows that the two luminosity distribution are different
at 92 per cent confidence ($\chi^2/\nu=14.2/8$), with the NLS1's have
systematically higher inferred $L/L_{Edd}$ than the non-NLS1 AGN, as
noted by e.g. Boroson (2002).

\begin{figure*}
\begin{center}
\begin{tabular}{l}
\leavevmode \epsfxsize=8.5cm \epsfbox{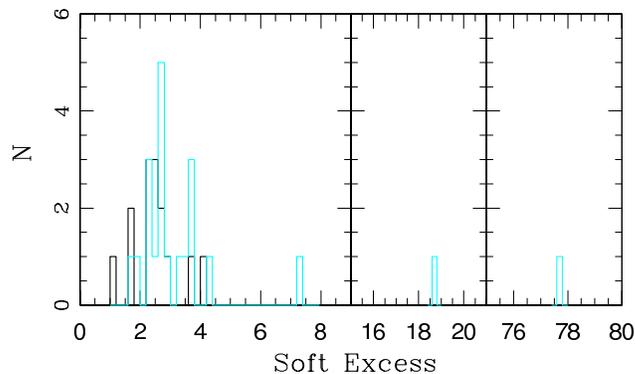}
\end{tabular}
\end{center}
\caption{Histogram showing the distribution of Soft Excess (SX)
at 0.5keV across the sample. The largest values of SX are 
found in the NLS1's (cyan) but there is also substantial overlap 
with the broad line AGN (black) at lower SXs.}
\label{fig:hist_softXS}
\end{figure*}

We follow Sobolewska \& Done (2007) and parameterise the strength of
the soft excess in each source by fitting a power law with galactic
absorption to the 2-10~keV spectrum and then taking the ratio of this
to the data at 0.5 keV. Fig.~\ref{fig:hist_softXS} shows the histogram
of these values. Whilst the NLS1's show the highest soft excess values the 
distribution is not significantly different to that of the broad line AGN. 
More than 70 per cent of the sample have soft excesses greater
than the maximum value from isotropic reflection of $\sim 2$ (see Fig.~\ref{fig:softXS_ag210}), and
three sources (PG1404+226, PHL 1092 and REJ1034+396) have soft
excesses $\ge 5$.

\begin{figure}
\begin{center}
\leavevmode \epsfxsize=7cm \epsfbox{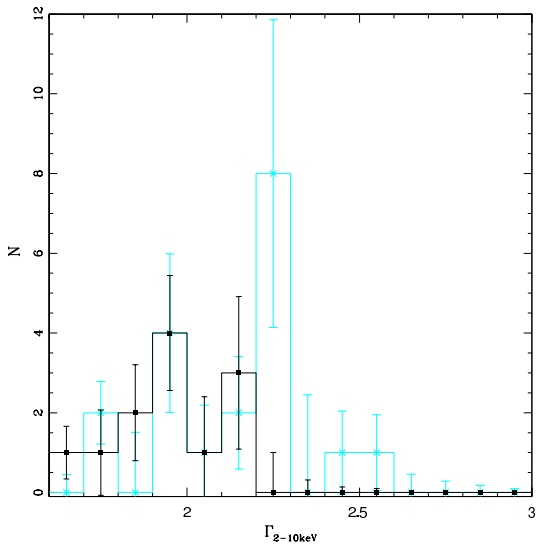}
\end{center}
\caption{Histogram of the 2--10~keV power law spectral index
for the NLS1's (cyan) and broad line AGN (black).
Error bars on the distribution are derived from a bootstrap Monte--Carlo
from the observed uncertainties on $\Gamma_{2-10~keV}$. 
Although the steepest spectral indices are associated with the NLS1s, 
the is substantial overlap between the two AGN types, and both show 
spectra significantly flatter than $\Gamma_{2-10~keV}=2$, which is 
very unusual in high mass accretion rate Black Hole Binaries.}
\label{fig:hist_ag210}
\end{figure}

\begin{figure}
\begin{center}
\leavevmode \epsfxsize=7cm \epsfbox{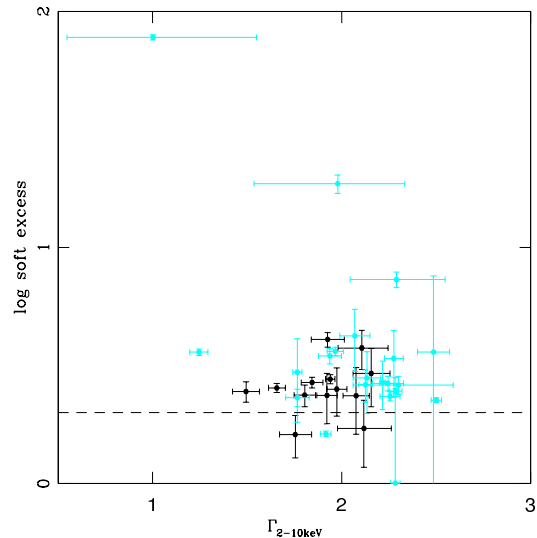}
\end{center}
\caption{Soft Excess strength (measured at 0.5keV) versus apparent
2--10~keV power law spectral index for NLS1's (cyan) and broad line AGN
(black). The object with the largest soft excesses also has a flat
spectrum, but in general there is no clear correlation, nor dependence
on AGN type. The dashed horizontal line shows the limit of soft excess that can be produced by a reflection model.}
\label{fig:softXS_ag210}
\end{figure}

Fig.~\ref{fig:hist_ag210} shows the distribution of apparent 2--10~keV
power law index.  For this, and for all other histograms of fit
parameters where the uncertainties are well defined, we use a
bootstrap Monte Carlo to assess the impact of these on each bin of the
histogram. The uncertainties (shown as an error bar on each bin) are
not independent as they have to sum to the given number of objects,
but it is clear that the NLS1s show a significantly different
distribution to the broad line AGN. However, this is not a simple
difference in mean spectral index as seen in previous studies where
NLS1's were significantly steeper than broad line AGN in 2--10~keV
(Brandt, Mathur \& Elvis 1997). Fig.~\ref{fig:hist_ag210} shows that
there are some NLS1 which are extremely flat, as well as some being
very steep. The two sets of objects have similar mean indices in these
samples, but the NLS1's have larger dispersion.  As pointed out by
Gierli'nski \& Done (2004), intrinsic power laws with $\Gamma < 2$ are
not compatible with the {\em steep} tail seen in similarly high mass
accretion rate BHB. Either the underlying emission mechanisms are
substantially different between the stellar and supermassive black
holes, or these spectra are highly distorted.

Fig.~\ref{fig:softXS_ag210} shows the soft excess strength versus the
2--10~keV power law spectral slope. This shows that there is no
overall correlation between these two. This is somewhat surprising, as
both reflection and absorption models would be expected to show such a
correlation in their simplest versions where the ionization state of
the material remains constant. However, observations show that the
variability is more complex. Detailed variability
studies of individual objects show that the spectra are steep at high
flux levels, with iron features in emission and a moderate soft
excess, while at minimum flux the 2--10~keV spectrum is very flat,
with an extremely strong soft excess and complex features at iron
(e.g. Miller et al 2007).

\section{Spectral fitting and results}

We now fit the full smeared absorption model, {\sc swind1} as
described by Gierli\'nski \& Done (2006) to the full 0.3--10~keV
spectral data. This is parameterised by the velocity dispersion
($\sigma$, with upper limit of $0.5c$ as the terminal velocity in an outflow is at
least twice the Gaussian dispersion), 
which controls the amount of smearing, as well as column
density, $N_H$, and ionization state $\xi$.  This smeared absorption
is applied to an intrinsic power law, and its reflected emission
(using the {\sc xion} hydrostatic disc models, with inner and outer
radii set to 30 and 1000$R_g$, respectively, leaving the solid angle
and ratio of X--ray to intrinsic disc illumination as free
parameters).  This total spectrum is further modified by galactic
absorption ({\sc wabs}, with minimum value set to the Galactic column
along the line of sight but with upper limit free to allow for host
galaxy absorption) and one or more warm absorbers ({\sc xstar} table
models 19 and 25, publicly available from the {\sc xstar} web
site\footnotemark
\footnotetext{ftp://legacy.gsfc.nasa.gov/software/plasma\_codes/xstar/xspectables},
parameterised by column density, $N_{H,XSTAR}$, and ionization state
$\xi_{XSTAR}$. We fix the redshift of both the narrow and smeared ionized absorbers to the
rest-frame of the AGN. Our model then allows enough free-parameters to
describe the data consistently across a broad sample and be well
constrained even in the case of poor sinal-to-noise.

\begin{table*}
\begin{minipage}{170mm} 

\bigskip

\begin{tabular}{cl|l|l|l|l|l|l}

  \hline
  \ Source &  $\chi^{2}_{\nu}$ (d.o.f) & $\chi^{2}_{\nu}$
  (d.o.f) & {\sc swind1} $\xi$ & {\sc swind1} $N_{H}$ & {\sc swind1} $\sigma$  & Intrinsic $\Gamma$
  & L$_{disc}$/L$_{bol}$\\
  & Reflection & Absorption &&&&\\
  \hline

 PG 0003+199$^{*}$  & 0.939 (968)& 1.214 (2112)& 2.879 $_{-0.035}^{+0.043}$& 11.1 $_{-1.2}^{+1.0}$ & 0.500  $_{-0.005}^{+peg}$   & 2.559  $_{-0.015}^{+0.014}$&0.409\\ 
 PG 0050+124$^{*}$ & 1.117 (924)& 1.183 (1616)& 3.608 $_{-0.047}^{+0.022}$ & 50.0 $_{-5.0}^{+peg}$& 0.500  $_{-0.033}^{+peg}$& 2.457 $_{-0.015}^{+0.013}$&0.518\\ 
 PG 0157+001  & 0.861 (234)& 1.030 (511)& 3.229 $_{-0.496}^{+0.460}$ & 19.4 $_{-11.8}^{+30.6}$ & 0.500  $_{-0.159}^{+peg}$& 2.379 $_{-0.114}^{+0.133}$&0.850\\   
 PG 0844+349$^{*}$   & 0.987 (627)& 1.039 (1361)& 3.075 $_{-0.070}^{+0.061}$ & 17.5  $_{-2.8}^{+3.1}$ & 0.333  $_{-0.026}^{+0.052}$& 2.513 $_{-0.021}^{+0.035}$&0.521\\  
 PG 0947+396  & 0.893 (477)& 0.944 (874)& 3.552 $_{-0.201}^{+0.087}$ & 50.0 $_{-22.2}^{+peg}$ & 0.500  $_{-0.141}^{+peg}$& 2.176 $_{-0.030}^{+0.044}$&0.865$^{OM}$\\  
 PG 0953+414  & 1.055 (512)& 1.065 (899)& 3.318 $_{-0.098}^{+0.158}$ & 34.9 $_{-8.1}^{+15.1}$ & 0.500   $_{-0.068}^{+peg}$& 2.349 $_{-0.045}^{+0.065}$&0.878$^{OM}$\\  
 PG 1048+342  & 1.043 (534)& 1.075 (907)& 3.103 $_{-0.087}^{+0.082}$ & 20.0 $_{-3.5}^{+12.8}$ &  0.500  $_{-0.197}^{+peg}$& 2.190 $_{-0.042}^{+0.032}$&0.914$^{OM}$\\  
 PG 1115+407$^{*}$& 0.913 (413)& 0.912 (746)& 3.597 $_{-0.137}^{+0.202}$ & 34.4 $_{-9.3}^{+15.6}$ &  0.482  $_{-0.110}^{+peg}$& 2.535 $_{-0.055}^{+0.059}$&0.945$^{OM}$\\  
 PG 1116+215  & 1.007 (397)& 0.900 (845)& 3.280 $_{-0.189}^{+0.240}$ & 25.4 $_{-8.0}^{+20.9}$ &  0.500  $_{-0.131}^{+peg}$& 2.406 $_{-0.045}^{+0.064}$&0.935$^{OM}$\\  
 PG 1202+281  & 1.009 (553)& 1.038 (992)& 3.174 $_{-0.093}^{+0.116}$ & 28.7 $_{-5.0}^{+4.1}$ &  0.445  $_{-0.081}^{+peg}$& 2.111 $_{-0.061}^{+0.046}$&0.810$^{OM}$\\  
 PG 1211+143$^{*}$   & 1.274 (936)& 1.400 (1608)& 2.547 $_{-0.006}^{+0.006}$ & 9.5 $_{-0.3}^{+0.2}$ &  0.114  $_{-0.004}^{+0.005}$& 2.556 $_{-0.010}^{+0.015}$&0.964$^{OM}$\\ 
 PG 1244+026$^{*}$   & 1.102 (292)& 1.070 (745)& 3.364 $_{-0.053}^{+0.022}$ & 50.0 $_{-7.0}^{+peg}$ &   0.500 $_{-0.016}^{+peg}$& 2.733 $_{-0.028}^{+0.031}$&0.306\\  
 PG 1307+085  & 0.871 (265)& 0.972 (608)& 3.007 $_{-0.131}^{+0.069}$ & 21.0 $_{-6.7}^{+5.8}$ &   0.358  $_{-0.072}^{+peg}$& 2.115 $_{-0.083}^{+0.094}$&0.971$^{OM}$\\  
 PG 1309+355  & 0.927 (293)& 0.972 (580)& 3.741 $_{-0.762}^{+0.173}$ & 34.8 $_{-30.8}^{+peg}$ &   0.153  $_{-0.105}^{+0.150}$& 2.030 $_{-0.038}^{+0.078}$&0.974$^{OM}$\\  
 PG 1322+659  & 1.009 (331)& 0.890 (614)& 3.050 $_{-0.189}^{+0.151}$ & 20.1 $_{-6.7}^{+9.0}$ &   0.413  $_{-0.118}^{+peg}$& 2.635 $_{-0.073}^{+0.081}$&0.901$^{OM}$\\  
 PG 1352+183  & 1.002 (370)& 1.070 (855)& 3.358 $_{-0.121}^{+0.066}$ & 49.8 $_{-12.9}^{+0.2}$ &  0.500  $_{-0.143}^{+peg}$    & 2.347 $_{-0.034}^{+0.078}$&0.734$^{OM}$\\  
 PG 1402+261$^{*}$ & 1.129 (384)& 0.901 (788)& 3.385 $_{-0.121}^{+0.064}$ & 45.4 $_{-13.1}^{+4.6}$ &  0.500   $_{-0.063}^{+peg}$& 2.441 $_{-0.034}^{+0.029}$&0.953$^{OM}$\\  
 PG 1404+226$^{*}$   & 0.893 (190)& 1.070 (364)& 2.953 $_{-0.061}^{+0.032}$ & 45.8 $_{-4.6}^{+3.4}$ &   0.346   $_{-0.045}^{+0.040}$& 2.880 $_{-0.067}^{+0.078}$&0.755\\  
 PG 1427+480  & 0.985 (524)& 1.127 (915)& 3.301 $_{-0.129}^{+0.225}$ & 27.6 $_{-5.9}^{+8.6}$ &   0.500  $_{-0.118}^{+peg}$& 2.245 $_{-0.081}^{+0.045}$&0.925$^{OM}$\\  
 PG 1440+356$^{*}$   & 0.989 (289)& 1.064 (1321)& 3.145 $_{-0.064}^{+0.088}$ & 26.4 $_{-4.9}^{+4.8}$ & 0.500 $_{-0.069}^{+peg}$& 2.701 $_{-0.028}^{+0.025}$&0.941$^{OM}$\\ 
 PG 1444+407  & 1.297 (289)& 1.156 (470)& 3.444 $_{-0.185}^{+0.264}$ & 49.9 $_{-22.1}^{+peg}$ &   0.500     $_{-0.210}^{+peg}$& 2.541 $_{-0.052}^{+0.064}$&0.937$^{OM}$\\  
 PG 1501+106  & 0.981 (931)& 1.250 (1655)& 3.086 $_{-0.021}^{+0.061}$ & 18.5  $_{-1.3}^{+3.9}$ &   0.332     $_{-0.026}^{+0.027}$& 2.381 $_{-0.029}^{+0.019}$&0.697\\ 
 NGC 4051$^{*}$ & 1.240 (1040)& 0.982 (597)& 2.817 $_{-0.060}^{+0.060}$ & 18.2 $_{-3.1}^{+2.8}$ &  0.375   $_{-0.083}^{+peg}$& 2.030 $_{-0.028}^{+0.037}$&0.729$^{BF}$\\ 
 IRAS 13349+2438$^{*}$  & 1.198 (664)& 1.372 (772)& 2.549 $_{-0.024}^{+0.026}$ & 10.9 $_{-0.8}^{+0.8}$ &  0.290  $_{-0.022}^{+0.025}$& 2.779 $_{-0.031}^{+0.032}$&0.917\\ 
ARK 564$^{*}$  & 1.197 (666)& 1.436 (1342)&3.406  $_{-0.041}^{+0.074}$ & 36.2 $_{-2.8}^{+12.1}$ &   0.500  $_{-0.006}^{+peg}$& 2.715 $_{-0.012}^{+0.011}$&0.877$^{OM}$\\
 MRK 1044$^{*}$ & 1.081 (530)& 1.144 (717)& 2.826 $_{-0.102}^{+0.075}$ & 12.6 $_{-2.1}^{+2.3}$ &   0.314  $_{-0.048}^{+0.170}$& 2.492 $_{-0.043}^{+0.039}$&0.593\\ 
 E 1346+266$^{*}$ & 1.068 (229)& 1.071 (343)& 3.372 $_{-0.171}^{+0.079}$ & 45.6 $_{-26.2}^{+peg}$ &   0.500 $_{-0.066}^{+peg}$& 2.900 $_{-0.050}^{+0.122}$&0.525$^{OM}$\\ 
 MRK 0359$^{*}$& 1.117 (527)& 0.983 (673)& 3.295 $_{-0.117}^{+0.128}$ & 9.3 $_{-0.4}^{+10.9}$ &  0.500   $_{-0.103}^{+peg}$& 2.083 $_{-0.036}^{+0.046}$&0.941\\ 
 PHL 1092$^{*}$ & 1.237 (180)& 1.279 (222)& 2.990 $_{-0.033}^{+0.025}$ & 50.0 $_{-2.8}^{+peg}$ &   0.500 $_{-0.024}^{+peg}$& 3.388 $_{-0.062}^{+0.066}$&0.741\\
 RE J1034+396$^{*}$  & 1.081 (282)& 1.105 (370)& 2.650 $_{-0.015}^{+0.046}$ & 16.7 $_{-1.5}^{+2.9}$ &   0.500  $_{-0.045}^{+peg}$& 3.357 $_{-0.054}^{+0.033}$&0.629$^{OM}$\\ 
 PKS 0558-504$^{*}$  & 1.063 (371)& 0.877 (949)& 2.751 $_{-0.071}^{+0.081}$ & 9.4 $_{-0.9}^{+1.8}$ &   0.500  $_{-0.017}^{+peg}$& 2.547 $_{-0.017}^{+0.017}$&0.932$^{OM}$\\
 MRK 0766 $^{*}$ & 1.119 (952)& 1.163 (760)& 2.661  $_{-0.042}^{+0.119} $ & 4.6 $_{-0.4}^{+0.8}$ &   0.277     $_{-0.025}^{+0.029}$& 2.222  $_{-0.010}^{+0.011}$&0.914$^{BF}$\\ 
 MRK 0586$^{*}$ & 1.007 (523)& 1.156 (524)& 2.861 $_{-0.167}^{+0.132}$ & 14.9 $_{-4.1}^{+2.1}$ &   0.351     $_{-0.076}^{+peg}$& 2.698 $_{-0.041}^{+0.051}$&0.952$^{OM}$\\ 
 TON S180$^{*}$  & 1.396 (778)& 1.207 (990)& 3.168 $_{-0.010}^{+0.024}$ & 24.1  $_{-1.3}^{+2.2}$ &  0.500      $_{-0.024}^{+peg}$& 2.632 $_{-0.009}^{+0.006}$&0.465\\ 

 \hline

\end{tabular}

\end{minipage}
\caption{Comparison of fit quality obtained through photoionized
reflection by Crummy et al. (2006) and the fits obtained here through
photoionized absorption. $^{*}$ corresponds to the source being a
NLS1. $^{peg}$ denotes that the upper or lower limit 
is outside of the tabulated parameter range.
Both models can adequately describe the range of spectra
seen in this heterogeneous sample of objects. The major parameters of
the smeared wind model (ionization parameter and column density)
are also tabulated (see Figs 5 and 6). 
L$_{disc}$/L$_{bol}$ is 
derived from simultaneous OM data where present (marked with $^{OM}$),
otherwise from disc fits to the archival optical data in Table 1.}
\label{tab:compare}
\end{table*}

\begin{figure}
\begin{center}
\leavevmode \epsfxsize=7cm \epsfbox{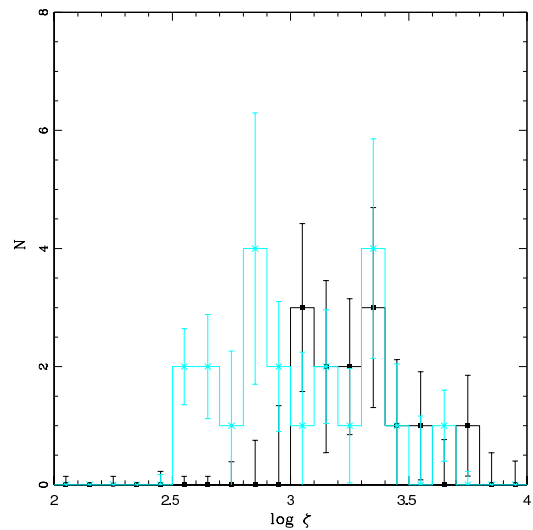}
\end{center}
\caption{Histogram showing the distribution of best-fit wind
ionization parameter with NLS1s shown in cyan and AGN in black.  
Clearly these are strongly peaked around
$\log\xi\sim 3$, as required in order to produce the soft excess from
the strong opacity jump from OVII/VIII. This fine tuning is rather
naturally produced if the absorbing clouds are at approximately
constant pressure. }
\label{fig:hist_ion}
\end{figure}

\begin{figure}
\begin{center}
  \leavevmode \epsfxsize=7cm \epsfbox{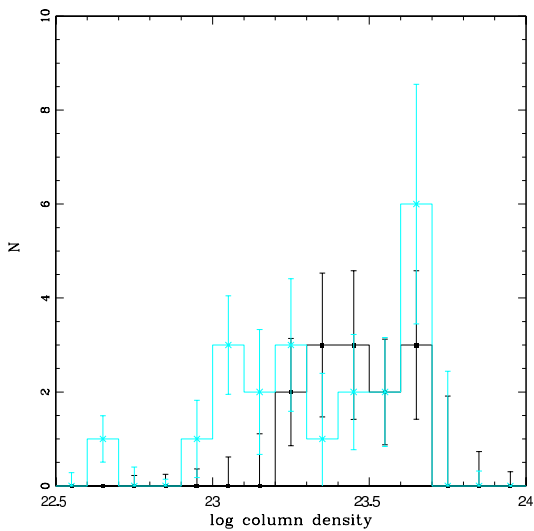}
\end{center}
\caption{Histogram showing the distribution of best-fit wind column
density. This is strongly peaked, showing again that a 
fine tuning mechanism is required. 
}\label{fig:hist_nh}
\end{figure}

The results of the fits with our absorption model are shown in
Table~\ref{tab:compare}. The reduced $\chi^2$ are statistically indistinguishable
from those obtained in the reflection fits of Crummy et al. (2006) over the large sample of objects. Both reflection and absorption
models have enough freedom to produce an acceptable fit the data over
the 0.3--10~keV bandpass of a wide range of different spectral shapes,
especially given the presence of the narrow warm absorber systems as
well.

Figs.~\ref{fig:hist_ion} and \ref{fig:hist_nh} shows the derived $\xi$
and $N_H$ for the wind. Plainly they are strongly peaked at $\sim
10^3$ and $3\times 10^{23}$ cm$^{-2}$, respectively. This apparent
'fine--tuning' can be explained as a consequence of the rapid change
in ionization which is produced self--consistently in X--ray
illuminated systems with some pressure balance condition (Krolik,
McKee \& Tarter 1981; Chevallier et al 2006). The X--ray illuminated
face of the cloud is heated to the Compton temperature, so expands
and has low density. This means its ionization parameter, $\xi=L/nr^2$
is high, and the material is fully ionised and hence
invisible. Further into the cloud the X-ray heating is less effective
due to scattering in the upper layers, so the density increases in
order to maintain pressure.  This increases the importance of
bremsstrahlung cooling, pulling the temperature down further, and
increasing the density. Eventually $\xi$ drops below $\sim 10^{3.5}$
where not all species are ionised and bound--free and bound--bound
transitions dramatically increase the cooling. The density drops very
rapidly, which further decreases the ionization, leading to the
possibility of yet more line transitions.  This runaway cooling only stops
when the material is nearly neutral, so a further decrease in
ionization does not lead to an increase in the available line
transitions. The corresponding rapid contraction of the cloud
(required to enhance the density) may lead to fragmentation of this
low ionization part of the cloud into clumps. A line of sight through
the cloud goes through the highly ionized layer (small opacity), the
transition layer (mean $\log\xi\sim 3$ and column of $\sim 10^{22-23}$
cm$^{-2}$) but is not likely to intercept any of the opaque cool
clumps as these have low covering factor (Chevallier et al 2006).

\begin{figure}
\begin{center}
\leavevmode \epsfxsize=7cm \epsfbox{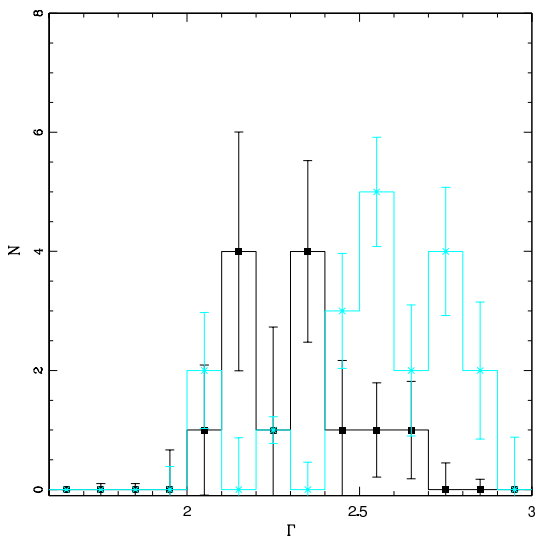}
\end{center}
\caption{Histogram showing the distribution of best-fit intrinsic
power law spectral index for the NLS1s (cyan) 
and broad line AGN (black). Unlike the apparent 2--10~keV spectral
index shown in Fig. 3, here there is a clear distinction between the
steeper NLS1 and flatter broad line AGN, and none of these have 
$\Gamma<2$ as expected by analogy with the high mass accretion rate
black hole binaries. 
}\label{fig:hist_gamma}
\end{figure}

Table 3 shows the individual best-fit wind velocity dispersions within
our model. The values are large, with many pegged at the upper limit
of $0.5c$. Most of these can be somewhat reduced allowing the smeared
absorber to be blue (or red) shifted. This has no effect on the
intrinsic spectral index but tends to decrease the derived column and
ionisation parameter by a factor 2 (fairly negligible on a log plot
such as Figs 6 and 7). Nonetheless, even with this extra degree of
freedom the velocity dispersion rarely drops below $0.2c$ due to the
requirement that the soft excess is a single smooth component, while
atomic processes give {\em two} strong features, associated with
OVII/VIII at 0.7-0.8 keV and Fe L at 1.5-2 keV, respectively. This is
still too large to be associated with the UV line driven disc wind
which {\em must} be present in such high $L/L_{Edd}$ AGN. Proga (2004)
show that such winds have terminal velocities of $\sim 0.2c$,
corresponding to a Gaussian dispersion at least a factor 2 smaller
(more precisely, $\sqrt{12}$ for constant acceleration: Schurch \&
Done 2007).  Thus this requires that the outflow is magnetically
driven (associated with the jet?) if the soft excess really is due to
smeared absorption.

Fig.~\ref{fig:hist_gamma} shows the inferred intrinsic power law
index. Unlike the apparent 2--10~keV index shown in Fig 3., there are
now {\em no} objects with spectra significantly harder than
$\Gamma=2$. This is not the case for the reflection models of Crummy
et al (2006), where PG~1048+342, PG~1211+143, PG~1307+085 and
(especially) Mkn~359 are all inferred to have hard spectra. This is an
important difference arguing in favour of the absorption modelling as
high mass accretion rate BHB almost always show soft spectra, so by
analogy we expect the high mass accretion rate AGN to do likewise.
This histogram also shows that the NLS1's have somewhat softer
spectra than the broad line AGN, with $<\Gamma >\sim 2.6$ as opposed
to $\sim 2.3$ (see also Brandt et al 1997). 
By analogy with the BHB, this might indicate that
the NLS1's correspond to the 'Very High State', characterised by a
soft tail, with $\Gamma\sim 2.5$, while the broad line AGN correspond
to the 'High/Soft' (or thermal dominant) state, where the tail has
$\Gamma\sim 2-2.2$ (Remillard \& McClintock 2006).

\begin{figure}
\begin{center}
\leavevmode \epsfxsize=8cm \epsfbox{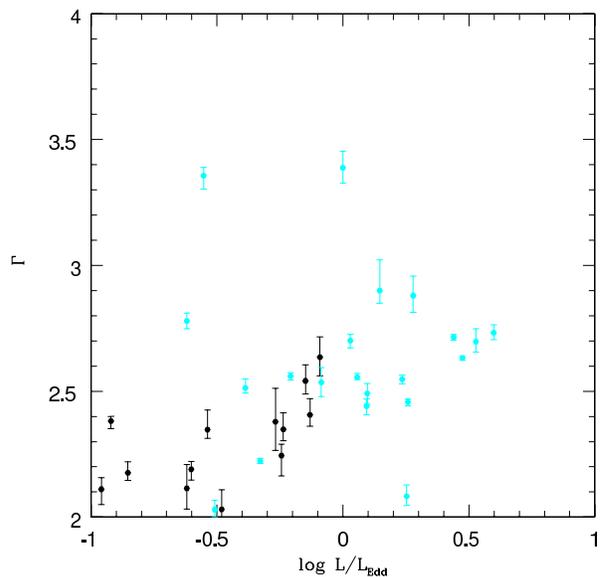}
\end{center}
\caption{Intrinsic power law index as derived from the absorption model versus
$L/L_{Edd}$. There is a clear correlation, especially given the
uncertainties in mass determination.}\label{fig:Gamma_Lall}
\end{figure}

Given that the NLS1 also have significantly higher $L/L_{Edd}$ than the broad 
line AGN (Fig.~\ref{fig:hist_Lall}) then it is also clear that in our sample the 
intrinsic spectral index will correlate with $L/L_{Edd}$. This is shown in 
Fig.~\ref{fig:Gamma_Lall}, and is remarkably well defined, especially 
considering the uncertainties in determining mass for these AGN. There is no 
such clear correlation in the Galactic binaries but this is probably due to 
hysteresis complicating their observed behaviour (e.g. Remillard \& McClintock 
2006). It is now clear that hysteresis in black hole binaries is connected to 
the dramatic disc instability (Gladstone, Done \& Gierlinski 2007), which 
probably does not occur in AGN (Done \& Gierlinski 2005). Thus it may be that 
AGN show the intrinsic (steady state) accretion flow properties more clearly 
than black hole binaries, and have spectral index correlated with $L/L_{Edd}$ 
(Shemmer et al 2006).

\begin{figure*}
\begin{center}
\leavevmode \epsfxsize=16cm \epsfbox{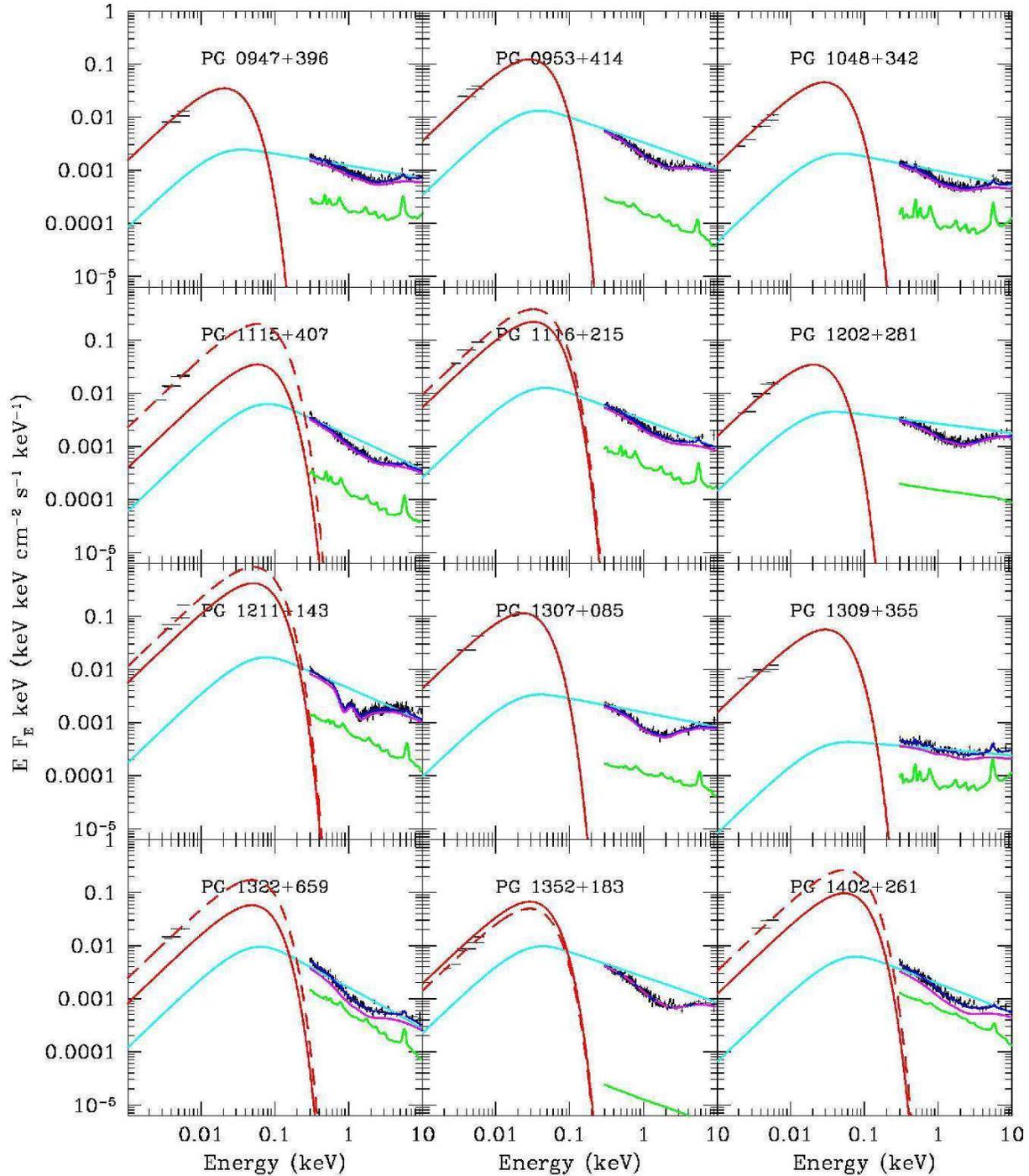}
\end{center}
\caption{Broadband $\nu f_\nu$ spectra of those sources with at least
2 XMM OM points. The data is absorption corrected, and decomposed into
disc (red), intrinsic Comptonized continuum (black), its reflection
(green) and the effect of the smeared wind on the continuum (magenta).
The total spectrum is given by the Compton continuum plus
reflection as absorbed by the smeared wind (blue). 
The disc flux was originally calculated from the archival tabulated
luminosity in Table 1 (red solid line). Since AGN vary, this can differ
substantially from the observed OM points, so the disc luminosity is
recalculated from the best fit to the OM points (red dashed line). 
}\label{fig:}
\end{figure*}

\begin{figure*}
\begin{center}
\leavevmode \epsfxsize=16cm \epsfbox{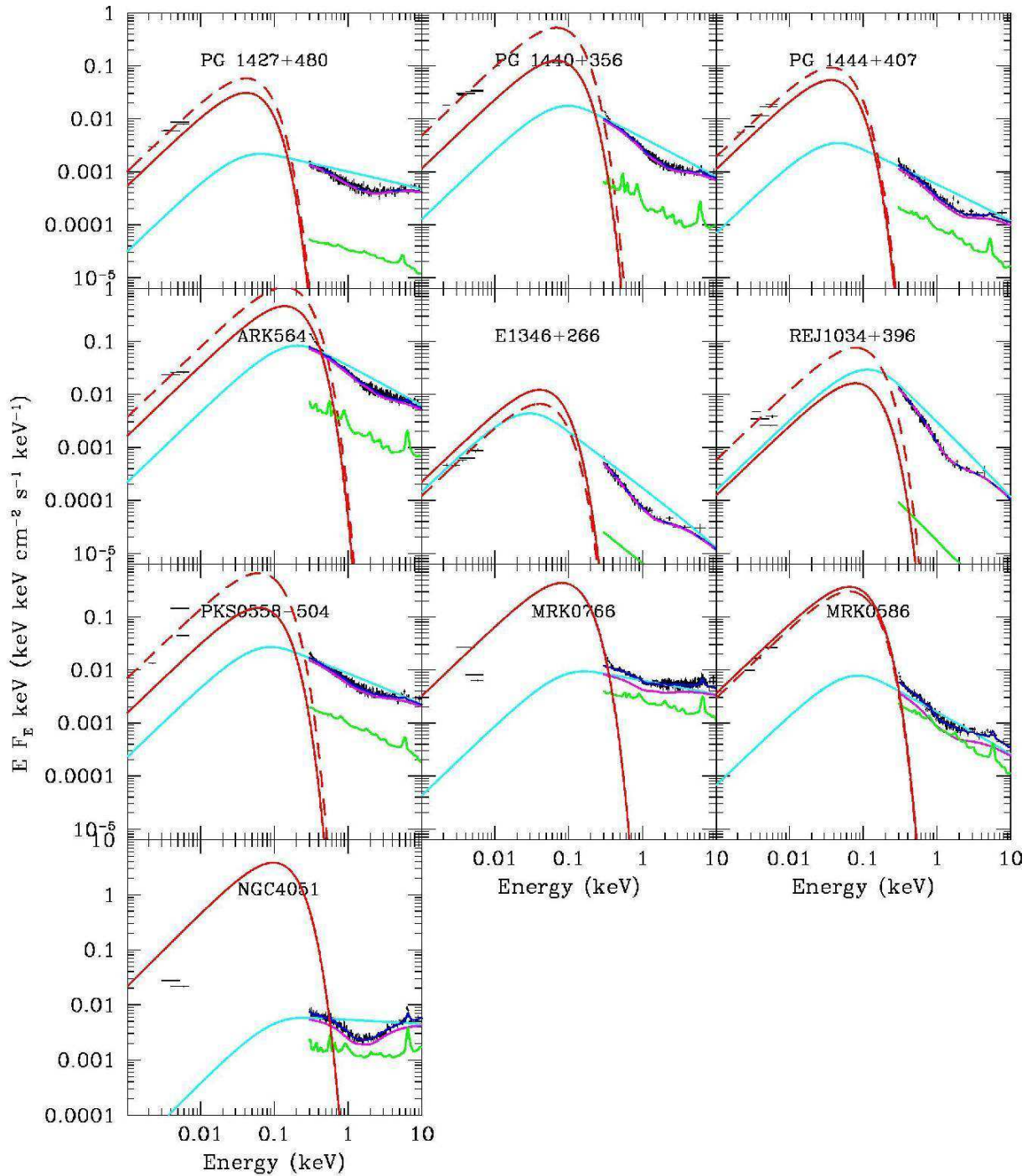}
\end{center}
\caption{As figure 9.}\label{fig:}
\end{figure*}

\begin{figure*}
\begin{center}
\leavevmode \epsfxsize=15cm \epsfbox{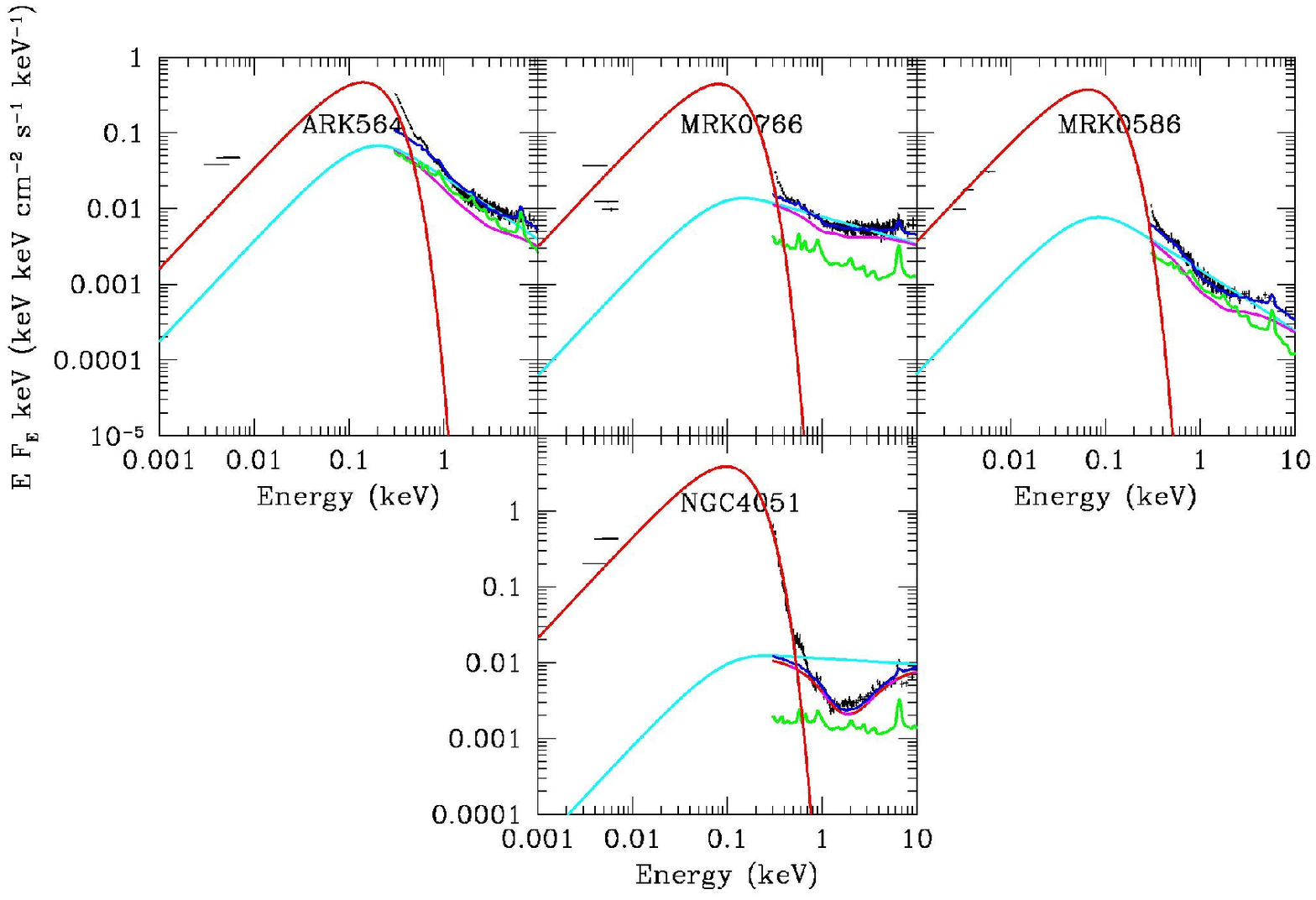}
\end{center}
\caption{The standard X-ray spectral model did not include the disc
component, but Fig 9 shows that there are some sources 
for which the disc temperature is high enough to contribute to the
soft X--ray flux. These sources also often also have OM points which are
significantly redder than expected for a disc. Refitting the X--ray data
including the disc emission requires a much higher neutral absorption column,
so giving a bigger correction for reddening, which can recover a
disc--like optical--UV spectrum. 
}
\label{fig:xraydisc}
\end{figure*}

\section{Simultaneous SED's}

The full observational criteria for BHB spectral state involve the
fraction of the disc to total luminosity as well as the shape of the
tail. For those observations for which there are at least two
XMM-\textit{Newton} OM data points we can construct the
$\nu f_\nu$ SED spanning the optical-UV-X--ray range,
i.e. covering both the disc and Comptonized tail. This means we can
characterise the full accretion power of the system, and derive the
disc to total power ratio, and hence determine spectral state. 

We use the parameters of our the best fitting model derived in Section
3 to deconvolve the data from the instrument response, and to correct
for neutral absorption (both from our Galaxy and the AGN host galaxy)
and narrow {\sc XSTAR} ionised absorption. On top of these absorption
corrected, deconvolved data, we also plot the best estimate for the
intrinsic Comptonized spectrum (cyan, with seed photons assumed to be
at the same temperature of the disc, see below) and the effect of
smeared absorption on this (magenta). The inferred intrinsic
reflection component is also included on the plot (green) and the
total (intrinsic plus reflection, both absorbed by the smeared wind)
fit to the data (blue).

We use the upper and lower limits on the (galactic and host Galaxy)
neutral $N_H$ to get upper and lower limits in the optical/UV
reddening (Predehl \& Schmidt 1995).  These are used to get
uncertainties on the dereddened optical/UV flux from the OM data.  We
include an additional standard {\sc diskbb} model with normalization
and temperature determined from the tabulated M$_{BH}$ and L/L$_{Edd}$
for each AGN (red). This should generally overestimate the optical/UV
flux, since this assumes that the entire bolometric luminosity is
produced in the disc. The fact that this generally matches to the data
shows immediately that this is a fairly good approximation for most of
these AGN, and that the high energy tail (cyan) is not energetically
dominant, though there are a few exceptions to this (E1346+266 and RE
J1034+396). Since the luminosity is taken simply from the tabulated
values then time variability is also an issue, hence we also show a
disc spectrum which gives a better match to the OM points (dotted red
line). 

There are a few objects where the optical/UV spectra have a very
different slope than that predicted by disc models, the NLS1's
MRK~766 and NGC~4051. These are also objects where the black hole
mass is low enough and $L/L_{Edd}$ high enough for the predicted disc
spectra to extend into the soft X--ray band. Since the disc was not
included in the X--ray spectral model then its presence would lead to
an underestimate of the X--ray column density. This would underestimate
the internal host galaxy reddening which could lead to the observed
downturn in optical/UV flux. Hence we refit the X--ray data of these objects
including the disc component. The increased reddening in NGC~4051
is sufficient to reverse the optical/UV slope, making it much more
compatible with the expected accretion disc spectrum. However, this is
not the case for MRK~766, where the optical data clearly appear absorbed
even with a larger host Galaxy column. We suggest that these data are
indeed strongly reddened, perhaps from a dust entrained in the narrow
warm absorber systems in a similar manner to MCG--6--30--15 (Reynolds
1997; Kraemer et al 2000; Ballantyne, Weingartner \& Murray 2003). 

\begin{figure}
\begin{center}
\leavevmode \epsfxsize=7cm \epsfbox{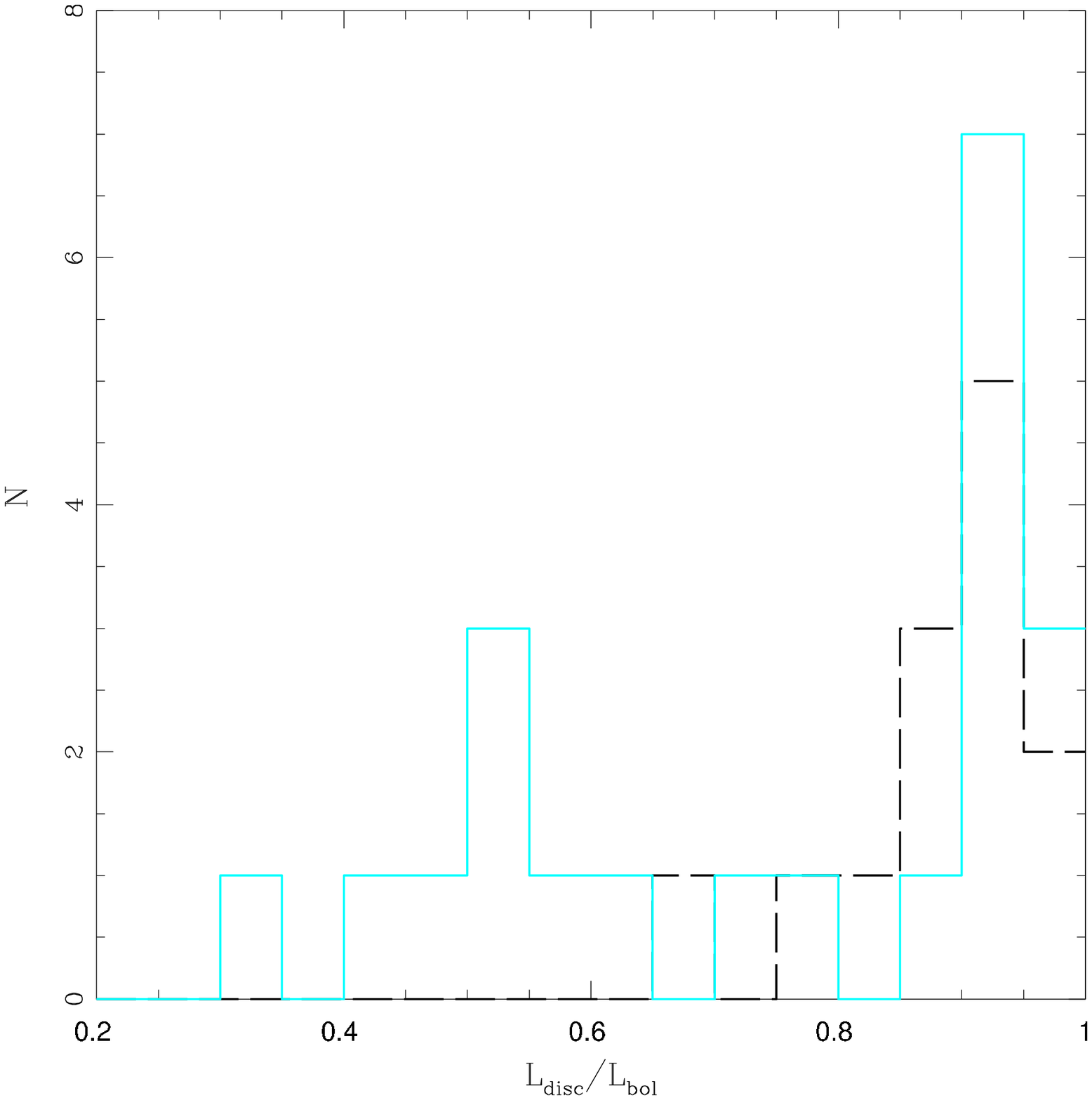}
\end{center}
\caption{Histogram showing the distribution of L$_{disc}$/L$_{bol}$.
While the spectral indices shown in Fig 7 allow an analogy between
NLS1's (cyan) with the very high (steep power law) state in BHB and
broad line AGN with the high/soft (thermal dominant) state, this is
{\em not} carried over into the ratio of disc to total luminosity. The
very high state is characterised by less than 80 per cent of power in
the disc, yet many NLS1's have estimated disc spectra which are more
dominant than this. 
}\label{fig:hist_Ldisc}
\end{figure}

We use these best estimates of the unabsorbed disc spectrum (dotted
red line) to quantify the ratio of luminosity in the tail to the total
luminosity or L$_{tail}$/L$_{tot}$. In BHB the High/Soft (thermal
dominant) State is defined as L$_{disc}$/L$_{tot} \ge$ 0.75, while the
Intermediate/Very High State (steep power law state) is defined to
have $L_{disc}/L_{tot}<0.8$ and $\Gamma>2.4$.  By contrast the
Low/Hard state has $L_{disc}/L_{tot}\le 0.2$ and $1.4<\Gamma < 2.1$.
The values of these ratios and the inferred analogous XRB states are
given in Table 6, and the histogram of this for the entire sample is
shown in Fig.~\ref{fig:hist_Ldisc}.

This shows a less clear cut distinction between the NLS1 and broad
line AGN than in Fig.~\ref{fig:hist_gamma}. While the NLS1 spectral
indices are indeed steep, as expected for the 'Very High State' in
BHB, they do {\em not} in general have $L_{disc}/L_{tot} < 0.8$. Also,
they do not in general have a significantly smaller $L_{disc}/L_{tot}$
than the broad line AGN (though the objects with the weakest discs are
NLS1's), as expected if the NLS1 correspond to the Very High State,
while the broad line AGN correspond to the High/Soft state. This lack
of correspondence with the BHB could be due to lack of data in the
extreme UV where the disc spectra peak. The disc luminosities are
inferred assuming that their spectra follow the expected standard
form, yet there is clear evidence in BHB that this is not the case in
the Very High State, where discs in BHB can show much lower maximum
temperature than expected for their luminosity (Kubota \& Done 2004;
Done \& Kubota 2006). If the NLS1's are indeed analogous to the Very
High state and have similarly distorted disc spectra then our estimate
of total disc luminosity from the optical/UV flux is too high. 
There is some evidence to support this in NGC~4051, as the high
temperature, luminous disc component shown in Fig.~\ref{fig:xraydisc}
is strongly ruled out by the XMM-Newton data. 

\section{Conclusions}

We show that the X--ray spectra of a heterogeneous sample of QSOs and
NLS1's can be equally well fit by partially ionised, Doppler smeared
material seen in absorption as by the the alternative geometry where
this material is optically thick and seen in reflection (Crummy et al
2006). Thus the 0.3--10~keV spectral data alone cannot distinguish the
underlying geometry of this partially ionised material, nor in fact
can it unambiguously show that it does indeed exist as the Doppler
smearing is so extreme as to merge the characteristic atomic features
into a quasi--continuum. Thus we have only indirect evidence for its
existence, namely the peak in rms variability spectra between
0.7--3~keV (e.g. Ponti et al 2006; Gierli\'nski \& Done 2006), and the
constant 'temperature' of the soft excess component, together with the
fact that this material can also distort the higher energy spectra,
producing occasional apparently anomalously hard (by comparison with
high $L/L_{Edd}$ BHB) 2--10~keV spectra. Similarly indirect arguments
from physical plausibility support a geometry in which this material
is optically thin and seen in absorption (Chevallier et al 2006;
Sobolewska \& Done 2007; Done \& Nayakshin 2007). Such absorption is
likely to be very complex, clumpy, and time dependent, with multiple
velocity structures.  

Nonetheless, despite the difficulties in producing realistic models to
compare with the data, even the simple models used here can begin to
reconstruct the intrinsic X--ray emission. These reinforce the links
between supermassive and stellar black holes, as the inferred
intrinsic X--ray spectra are all soft, as expected by analogy with the
high mass accretion rates (high/soft and very high) states in the
binary systems.  All objects with the apparently hard 2--10~keV
spectral indices can be explained as being intrinsically soft, but
strongly distorted by the smeared absorption.  Furthermore, there is a
clear correlation between the intrinsic X--ray index and the type of
AGN, with NLS1's having softer X--ray spectra than broad line AGN,
consistent with them being the analogues of the Very High State while
the broad line AGN are identified with the High/Soft State.
respectively. However, this analogy does {\em not} appear to hold in
terms of the ratio of disc to total luminosity. The Very High state in
BHB has the disc be less than 80 per cent of the total accretion
power, while the NLS1's show a range of values, with many having the
disc more dominant than expected. This could point to subtle
differences in how spectral states are seen between stellar and
supermassive black holes, but we stress that this could also result
from a lack of data in the EUV region where the disc luminosity peaks.
Disc spectra in the Very High State in BHB are often somewhat
distorted. Similar distortions in the NLS1 would result in an
overestimate of the total disc flux inferred from the observed
optical/UV points. This could recover a straightforward correspondence
between the spectral states as seen in BHB and AGN type. 

\label{lastpage}

\end{document}